\documentclass[12pt]{article}
\usepackage{amsmath,amssymb}    
\usepackage{amsfonts}
\usepackage{enumerate}
\usepackage{ytableau}
\usepackage{hyperref}
\usepackage{bm}
\usepackage{tikz}
\begin{document}
\def\be{\begin{equation}}
\def\bea{\begin{eqnarray}}
\def\ee{\end{equation}}
\def\eea{\end{eqnarray}}
\def\d{\partial}
\def\eps{\varepsilon}
\def\la{\lambda}
\def\b{\bigskip}
\def\nn{\nonumber \\}
\def\p{\partial}
\def\t{\tilde}
\def\h{{1\over 2}}
\def\be{\begin{equation}}
\def\bea{\begin{eqnarray}}
\def\ee{\end{equation}}
\def\eea{\end{eqnarray}}
\def\b{\bigskip}
\def\u{\uparrow}
\def \byt{\begin{ytableau}}
\def \eyt{\end{ytableau}}

\makeatletter
\def\blfootnote{\xdef\@thefnmark{}\@footnotetext}  
\makeatother


\title{\textbf{Holographic Correlators on Integrable Superstrata}}

\vspace{14mm}
\author{
Jia Tian$^{1,3}$, Jue Hou$^{1}$ and Bin Chen$^{1,2,3}$\footnote{bchen01, houjue, wukongjiaozi@pku.edu.cn}
}
\date{}

\maketitle

\begin{center}
{\it
$^{1}$Department of Physics and State Key Laboratory of Nuclear Physics and Technology,\\Peking University, 5 Yiheyuan Rd, Beijing 100871, P.~R.~China\\
\vspace{2mm}
$^{2}$Collaborative Innovation Center of Quantum Matter, 5 Yiheyuan Rd, Beijing 100871, P.~R.~China\\
$^{3}$Center for High Energy Physics, Peking University, 5 Yiheyuan Rd, Beijing 100871, P.~R.~China
}
\vspace{10mm}




\end{center}

\begin{abstract}

In this work, we study the $\frac{1}{8}$-BPS heavy-heavy-light-light correlators in the D1D5 CFT and its holographic dual. On the field theory side, we compute the fermionic four-point correlators at the free orbifold point. On the dual gravity side, we compute the correlators of the scalar operators in the supergravity limit of the D1D5 CFT. Following the strategy of \cite{Galliani:2017jlg}, the four-point function is converted into a two-point function in non-trivial geometries known as superstrata which are supergravity solutions preserving $1/8$ supersymmetries. We focus on a family of integrable superstrata, which allows us to compute the correlators perturbatively. 

\end{abstract}

\baselineskip 18pt

\newpage



\section{Introduction}
\label{Intorduction}
The analysis of four-point correlators provides important insights to the study of $AdS/CFT$ duality \cite{Maldacena:1997re,Gubser:1998bc,Witten:1998qj}. These correlators are not protected against renormalization group (RG) flow so that they encode non-trivial dynamical information of the theory. In the gravity side, the standard approach of calculating four-point correlators is to use Witten's diagrams. While in the context of $AdS_3/CFT_2$ duality, this approach suffers several technical difficulties \cite{DHoker:1999mqo} which render the derivation of exact four-point correlators challenging. Nevertheless, a subclass of four-point correlators denoted as HHLL which contain two heavy and two light operators has been studied both from the CFT and gravity points of view \cite{Balasubramanian:2005qu}--\cite{daCunha:2016crm}. The goal of this paper is to extend these results to other sector of this duality.

In the HHLL correlator we consider, the conformal dimension of the heavy operators scales as the central charge $c$ of the theory when $c\rightarrow \infty$. This type of heavy operators has strong back-reactions on the background so they are dual to new asymptotically $AdS$ geometries. In contrast to the previous works where the heavy states are dual to black holes, the authors of \cite{Galliani:2017jlg} considered a special type of heavy operators which are dual to simple supergravity configurations known as the microstate geometries. These microstate geometries are horizonless and smooth and are proposed to describe the microstates of black holes. On the CFT side, they correspond to states in Ramond-Ramond (RR) sector which have conformal dimension of order $c$. The states considered in \cite{Galliani:2017jlg} belong to the $1/4$-BPS sector. In the paper we extend the analysis to the $1/8$-BPS sector where the heavy states can describe black holes with macroscopic horizons.

Some special correlators with $1/8$-BPS states have been studied in \cite{Galliani:2016cai}. But those states are not typical in the sense that their dual geometries are locally isometric to $AdS_3\times S^3$ so they do not contribute to the entropy of black holes.
In the $1/8$-BPS sector, the microstate geometries of  typical states were firstly constructed in \cite{Bena:2015bea} and more solutions were found  in \cite{Bena:2017xbt,Tian:2016ucg,Ceplak:2018pws,Bakhshaei:2018vux,Heidmann:2019zws}\footnote{For a non-supersymmetric example see \cite{Bombini:2017got}}. These microstate geometries which are called the superstrata in the D1-D5-P frame are asymptotic to $AdS_3\times S^3$ but have very complicated geometries near the horizon regions. Fortunately, one family of superstrata was found to possess integrable structure \cite{Bena:2017upb} which makes the holographic calculations of the four-point correlators feasible. In this paper, we present the holographic computations in this family of geometries. 

This paper has the following organization. In section \ref{CFT}, after a brief review of the D1-D5  CFT, we will compute the four-point HHLL correlators at free symmetric orbifold CFT point. The heavy states we consider are composed of two kinds of RR vacua with momentum excitations.  In section \ref{Gravity},  we first introduce the superstrata solutions which are dual to the heavy states we consider in the CFT and then compute the correlators perturbatively by taking the limit where the number of one kind of RR vacuum is much larger than the one of the other. Some technical details are presented in the appendices.

\textbf{Note added:} when this work was finalizing, we realized that the related work \cite{Bombini:2019vnc} derived the similar results. There they also computed the holographic $1/8$-BPS HHLL correlators in the same geometries as we considered.

\section{The CFT picture}
\label{CFT}
\renewcommand{\theequation}{2.\arabic{equation}}
\setcounter{equation}{0}
In this section we compute the four-point correlators in the D1D5 CFT at the orbifold point. As a setup we begin with the brief review of the orbifold CFT\footnote{See \cite{Moscato:2017usq} for more details.}. The target space of this CFT is $(T^4)^N/S_N$ so it can be formulated in terms of $N$ groups of free bosonic and fermionic fields
\bea \label{FieldCont}
\p X_{(r)}^{\dot{A}A}(z),\bar{\p}X_{(r)}^{\dot{A}A}(z),\psi_{(r)}^{\alpha \dot{A}}(z),\tilde{\psi}_{(r)}^{\dot{\alpha} \dot{A}}(\bar{z}),
\eea 
where $r=1,\dots,N$ runs over different groups for which we also call strands and  $\alpha,\dot{\alpha}=1,2$ are the spinorial indices for the R-symmetry group $SU(2)_L\times SU(2)_R$ while $(A,\dot{A})=1,2$ are indices for the $SU(2)_1\times SU(2)_2=SO(4)_I$ rotations acting on the tangent space in the compact manifold $T^4$. In this paper, we only consider the untwisted sector of this orbifold theory, on which the operators can be written as direct products of operators acting on each strand.

\subsection{Heavy and light operators}
The structure of the correlators that we consider is
\bea 
\langle O_H(z_1)\bar{O}_H(z_2)O_L(z_3)\bar{O}_L(z_4)\rangle=\frac{1}{z_{12}^{2h_H}z_{34}^{2h_L}}\frac{1}{\bar{z}_{12}^{2\bar{h}_H} \bar{z}_{34}^{2\bar{h}_L}}\mathcal{G}(z,\bar{z}),
\eea 
where $z_{jk}=z_j-z_k$ and $\mathcal{G}$ is a function of conformal cross ratios
\bea 
z=\frac{z_{14}z_{23}}{z_{13}z_{24}},\quad \bar{z}=\frac{\bar{z}_{14}\bar{z}_{23}}{\bar{z}_{13}\bar{z}_{24}}.
\eea 
The light operators we are interested in are chiral primaries \footnote{We adopt the notations used in \cite{Galliani:2017jlg}.}
\bea \label{LightOperator}
O_L=\frac{1}{\sqrt{N}}\sum\limits_{r=1}^N O^L_{(r)},\quad O^L_{(r)}=-\frac{i}{\sqrt{2}}\psi^{1\dot{A}}_{(r)}\epsilon_{\dot{A}\dot{B}}\tilde{\psi}^{\dot{1}\dot{B}}_{(r)},
\eea 
whose comformal dimension is $h_L=\bar{h}_L=1/2$. The heavy operators that are considered in \cite{Galliani:2017jlg} correspond to the Ramond-Ramond (RR) ground states
\bea 
O_H=\otimes _{r=1}^N O^H_{(r)},\quad O^H_{(r)}\equiv |m\rangle,
\eea 
where $m=(0,0),(\pm,\pm)$ could be one of the five possible spin states. Here $m$ is the quantum number of R-symmetry group $SU(2)_L\times SU(2)_R$ whose current operators read
\bea 
&& J^{\alpha\beta}(z)=\sum\limits_{r=1}^{N}J_{(r)}^{\alpha\beta},\quad \tilde{J}^{\dot{\alpha}\dot{\beta}}(z)=\sum\limits_{r=1}^{N}J_{(r)}^{\dot{\alpha}\dot{\beta}},\\
&&J_{(r)}^+=\frac{1}{2}\epsilon_{\dot{A}\dot{B}}\psi^{1\dot{A}}\psi^{1\dot{B}},~~ J_{(r)}^-=-\frac{1}{2}\epsilon_{\dot{A}\dot{B}}\psi^{2\dot{A}}\psi^{2\dot{B}},~~
J_{(r)}^3=-\frac{1}{2}(\epsilon_{\dot{A}\dot{B}}\psi^{1\dot{A}}\psi^{2\dot{B}}-1).\nonumber
\eea

The coherent superposition of states are dual to $1/4$-BPS geometries since the momentum charge $p=h_H-\bar{h}_H=0$. To extend to the $1/8$-BPS sector, one can either perform spectral flows \cite{Lunin:2012gp,Bena:2008wt} or add proper excitation  \cite{Giusto:2013rxa}. In the gravity side, the dual geometries of the former cases are locally isometric to $AdS_3$. While the later cases can have more interesting geometric duals, for example the microstate geometries of the black holes. In this paper we concentrate on such a set of $1/8$ BPS states of the form
\bea \label{State}
|m,p,n\rangle\equiv (|++\rangle)^{N-p}(\frac{(J_{-1}^+)^m}{m!}\frac{(L_{-1}-J_{-1}^3)^n}{n!}|00\rangle)^{p},
\eea 
where $L_{-1}$ is the expending mode of the Virasoro algebra. The operator $(L_{-1}-J_{-1}^3)^n$ increases the conformal dimension $h$ by $n$ so as to  add $n$ units of momentum charges. In order to have a smooth geometric dual one needs to take a coherent linear combination as
\bea \label{HeavyOperator}
|s_B\rangle=\frac{1}{N^{\frac{N}{2}}}\sum\limits_{p=0}^N A^{N-p}B^p |m,p,n\rangle,\quad |A|^2+|B|^2=N.
\eea 

In summary, we will compute the $1/8$-BPS correlators involved with (\ref{LightOperator}) and (\ref{HeavyOperator})
\bea\label{Correlator}
\langle s_B|O_L\bar{O}_L|s_B\rangle,
\eea 
where we have chosen $z_2\rightarrow \infty$ and $z_1\rightarrow 0$.

\subsection{Fermionic four-point  correlators}

In this subsection we first elaborate the calculation of the correlator (\ref{Correlator}) with $n=1$, and then present the results for generic $n$.
 When $n=1$ we need to compute 
 \bea\label{N1Correlaton}
 \frac{1}{N}\sum\limits_{r,s=1}^{N} \langle 0,p,1| O^L_{(r)}\bar{O}^L_{(s)} |0,p,1\rangle 
 \eea 
 with $|0,p,1\rangle$ given in \eqref{State}.
 Similar to the correlators in $1/4$ BPS sector \cite{Galliani:2017jlg}, this correlator in the large $N$ limit has two types of terms. The last term scales as $\mathcal{O}(N)$ due to the contributions from the terms where two light operators act on different strands, while the first two terms are of order $\mathcal{O}(1)$ due to the ``diagonal" contribution where light operators act on the same strand.

First let us compute the off-diagonal contribution. Because these two light operators act on different strand independently the four-point correlator splits into a product of two three-point functions
\bea 
\sum\limits_{r\ne s} \langle s_B | O_{(r)}^{1\dot{1}}(z_3) O_{(s)}^{2\dot{2}}(z_4) |s_B\rangle=\sum\limits_{r\ne s}\langle O_{}^{1\dot{1}}(z_3)(L_{-1}-J_{-1}^3)\rangle_{r} \langle (L_{-1}-J_{-1}^3)^\dagger O_{}^{2\dot{2}}(z_4)\rangle_{s}.
\eea 
Considering the fact that the zero mode of $O^{1\dot{1}}$ turns the state $|00\rangle$ into $|++\rangle$ and vice versa for $O^{2\dot{2}}$  the three point functions are given by
\bea 
\langle ++|O_{}^{1\dot{1}}(z_3)(L_{-1}-J_{-1}^3)|00\rangle 
&=&\langle++ |[O_{}^{1\dot{1}}(z_3),L_{-1}]|00\rangle+\langle++ |[J^{3}_{-1},O_{}^{1\dot{1}}(z_3)]|00\rangle \nonumber \\
&=& z_3^{-3/2}\bar{z}_3^{-1/2},\\
\langle 00|(L_{+1}-J_{+1}^3)O^{2\dot{2}}(z_4)|++\rangle &=&\langle 00|[L_{+1},O^{2\dot{2}}(z_4)]|++\rangle-\langle 00|[J_{+1},O^{2\dot{2}}(z_4)]|++\rangle \nonumber\\
&=&z_4^{1/2}\bar{z}_4^{-1/2}.
\eea 
Here we have used the commutation relations
\bea \label{Commutation}
&&[O(z),L_n]=-h_O (n+1)z^n O(z)-z^{n+1}\p_z O(z),\nonumber \\
&&[J_n^3,O^{1\dot{1}}(z)]=\frac{1}{2}z^nO^{1\dot{1}}(z),\quad [J_n^3,O^{2\dot{2}}(z)]=-\frac{1}{2}z^nO^{2\dot{2}}(z).
\eea 
By choosing $z_3=1$ and $z_4=z$, the off-diagonal contribution is
\bea 
\mathcal{G}_{off}(z,\bar{z})&=&z\frac{|1-z|^2}{|z|N}\sum\limits_{p=0}^Np(N-p) \frac{|A|^{2(N-p)}|B|^{2(p)}}{N^N}
\begin{pmatrix}
	N  \\
	 p \\
\end{pmatrix}
\nonumber\\
&=&z\frac{|1-z|^2}{|z|N}\frac{N(N-1)|A|^2|B|^2(|A|^2+|B|^2)^{N-2}}{N^N}\nonumber \\
&=&z\frac{|1-z|^2}{|z|}\frac{|A|^2|B|^2(N-1)}{N^2},
\eea 
where $\tiny \begin{pmatrix}
N  \\
p \\
\end{pmatrix}$ is the norm of $|++\rangle^{N-p}|00\rangle^p$, the extra $N$ comes from the normalization of $O_L$, and the factors  $p$ and $(N-p)$ follow from the action of $O^{1\dot{1}}$ and $O^{2\dot{2}}$ on the different copies respectively.

There are two building blocks we need to consider in the diagonal contributions
\bea 
&&\langle ++|O_L \bar{O}_L |++\rangle,  \\
&& \langle 00|(L_+-J_+ ^3)O_L \bar{O}_L (L_--J_- ^3)|00\rangle. 
\eea 
The first one does not depend on $n$ and it  has been computed in \cite{Galliani:2017jlg}:
\bea \label{DiagonalA}
\mathcal{G}_{diag}^A=\frac{1}{|z|}\frac{|A|^2}{N}.
\eea 
Let us focus on the second one. 
For simplicity, we define new operators $M_\pm$ as
\begin{equation}\label{}
M_\pm\equiv L_\pm-J^3_\pm,
\end{equation}
which satisfy the following useful commutation relations 
\bea\label{Commutation2}
&&	[M_+,M_-]=1+2L_0-2J_0, \\
&&	\left[ J _ { 0 } , M _ { \pm} \right]=0,\quad \left[ L _ { 0 } , M _ {\pm } \right]=\mp M _ {\pm }. 
\eea
Acting $M_\pm$ on the operator $\{O^{1\dot{1}}(z_3),O^{2\dot{2}}(z_4)\}\equiv \{O_3,O_4\}$, according to \eqref{Commutation2}, one can get
\bea
&&[M_{\pm},O_3]=D_{\pm}^3 O_3,\quad [M_{\pm},O_4]=D_{\pm}^4 O_4,\\
&&D_-^3\equiv\p_3-\frac{1}{2}z_3^{-1},\quad D_+^3\equiv\frac{1}{2}z_3+z_3^2\p_3,\quad D_-^4\equiv\p_4+\frac{1}{2}z_4^{-1},\quad D_+^4\equiv\frac{3}{2}z_4+z_4^2\p_4. \nonumber
\eea
Therefore the second diagonal block is equal to
\begin{equation}\label{}
\begin{split}
&\langle 00|(L_+-J_+ ^3)O_L \bar{O}_L (L_--J_- ^3)|00\rangle=\langle M_+O_3 O_4 M_-\rangle\\
=&\langle O_3 O_4 [M_+,M_-]\rangle+D_+\langle O_3 O_4 M_-\rangle=(1-D_+D_-)\langle O_3 O_4\rangle\\
\end{split}
\end{equation}
where $D_\pm\equiv D^3_\pm+D^4_\pm$. And we have used  $M_-|00\rangle=\langle 00|M_+=0$ ,  $L_0|00\rangle=J_0|00\rangle=\langle00|L_0=\langle00|J_0=0$\footnote{$L_0$ is defined in the R sector here}. Finally plugging the expression of $\langle O_3 O_4\rangle$
\bea \label{O34Expression}
\langle O_3 O_4\rangle=\langle 00|O_L\bar{O}_L|00\rangle&=&\frac{1}{2}\frac{1}{|z_3z_4||z_3-z_4|^2}(|z_3|^2+|z_4|^2+|z_3-z_4|^2),
\eea 
into the last equation and, one can obtain
\begin{equation}\label{}
\begin{split}
\langle 00|(L_+-J_+ ^3)O_L \bar{O}_L (L_--J_- ^3)|00\rangle = \frac{1+z+\bar{z}-z^2-2|z|^2(1-z)}{2|z||1-z|^2},
\end{split}
\end{equation}
which leads to the diagonal contribution 
\bea\label{GeneralOff}
\mathcal{G}^B_{diag}=\frac{1+z+\bar{z}-z^2-2|z|^2(1-z)}{2|z|} \frac{|B|^2}{N}.
\eea 
Summing $\mathcal{G}^{(A,B)}_{diag}$ and $\mathcal{G}_{off}$, we obtain 
\bea \label{Correlator1}
\mathcal{G}(z,\bar{z})=\frac{1}{|z|}+\frac{|B|^2}{2N}\frac{\bar{z}-1+z(z-1)(2\bar{z}-1)}{|z|}+\frac{|A|^2|B|^2}{N}(1-\frac{1}{N})\frac{z|1-z|^2}{|z|}.
\eea

For generic $n$, we present the expression of $\mathcal{G}_{off}$ and a recurrence equation of $\mathcal{G}^B_{diag}$ here and leave the derivation and the solution of recurrence equation to the Appendix \ref{GeneralN}. The off-diagonal contribution reads
\bea \label{OffDiagonalN}
\mathcal{G}_{off}(z,\bar{z})=z^n\frac{|1-z|^2}{|z|}\frac{|A|^2|B|^2(N-1)}{N^2}.
\eea 
Denoting the correlator $\langle 00|M_+^nO_3O_4 M_-^n|00\rangle$ by $G_n$, we get
\bea\label{GenralDiagonalA}
&&G_{n+2}=[(n+1)^2+(n+2)^2-D]G_{n+1}-(n+1)^4 G_{n},\\
&&G_{1}=(1-D)G_{0},\quad G_0=\langle O_3O_4\rangle,\quad D\equiv D_-D_+.\nonumber
\eea 
Therefore the corresponding diagonal contribution to the correlator is 
\bea 
\mathcal{G}^B_{diag}=\frac{G_n}{n!^2}\frac{|B|^2}{N}.
\eea 
Adding all the terms together gives the final expression 
\bea \label{CorrelatorN}
\mathcal{G}(z,\bar{z})=\frac{1}{|z|}+\frac{|B|^2}{N}\left(\frac{G_n}{n!^2}-\frac{1}{|z|}\right)+z^n\frac{|1-z|^2}{|z|}\frac{|A|^2|B|^2(N-1)}{N^2}.
\eea 
This expression manifests the feature of the $1/8$ BPS sectors that it is asymmetric with respect to $z$ and $\bar{z}$.

The correlators \eqref{Correlator} we have considered in this section are called fermionic correlators which are closed related to another type of correlators which are called bosonic correlators via the Ward identity \cite{Galliani:2017jlg} \footnote{This supersymmetric Ward identity does not hold exactly with the existence of $1/8$ BPS states. Thank Alessandro Bombini for pointing this our.  Here we made an assumption the symmetry were still approximately preserved under our perturbative calculation in next section. }
\bea\label{WardIdentity}
\langle \bar{O}_H|O_{bos}(1)\bar{O}_{bos}(z,\bar{z})|O_H\rangle=\p\bar{\p}[|z|\langle \bar{O}_H|O_{fer}(1)\bar{O}_{fer}(z,\bar{z})|O_H\rangle].
\eea 
The bosonic light operator in \eqref{WardIdentity} is a superdescendant of the chiral primary light operator \eqref{LightOperator} and it is defined as
\bea \label{BosonicLightOperator}
O_{bos}=\sum\limits_{r=1}^N \frac{\epsilon_{\dot{A}\dot{B}}}{\sqrt{2N}}\p X^{1\dot{A}}_{(r)}\bar{\p}X^{1\dot{B}}_{(r)},\quad \bar{O}_{bos}=\sum\limits_{r=1}^N \frac{\epsilon_{\dot{A}\dot{B}}}{\sqrt{2N}}\p X^{2\dot{A}}_{(r)}\bar{\p}X^{2\dot{B}}_{(r)}.
\eea 
Both the fermionic and bosonic operators can be identified with the marginal operators in the moduli space of the CFT and they are dual to different fluctuations on the geometric backgrounds sourced by the heavy operators. In particular,  it has been shown in \cite{Galliani:2017jlg} that the bosonic operators are described by minimally coupled scalars whose wave equations are rather simple comparing to the ones of other fluctuations. Hence in the next section, we will compute the bosonic correlators \eqref{WardIdentity} in the supergravity limit.

\section{The gravity picture}
\label{Gravity}
\renewcommand{\theequation}{3.\arabic{equation}}
\setcounter{equation}{0}
In this section we compute the four-point correlator \eqref{WardIdentity} holographically in the supergravity limit. We first introduce the  $1/8$-BPS background geometries known as the superstrata and then focus on an integrable family of the geometries which are dual to the states \eqref{HeavyOperator} in the CFT.

\subsection{$1/8$-BPS geometries: superstrata}
Superstrata are microstate geometries of five-dimensional, three charge, supersymmetric black holes with arbitrarily small angular momenta in the D1-D5-P frame. They are regular solutions of the six-dimensional truncation of type IIB supergravity on $M^{4,1}\times S^1$, whose metric take the general form \cite{Gutowski:2003rg}
\bea \label{6DMetrix}
ds_6^2&=&-\frac{2}{\sqrt{\mathcal{P}}}(dv+\beta)(du+\omega+\frac{\mathcal{F}}{2}(dv+\beta))+\sqrt{\mathcal{P}} ds_4^2(\mathcal{B}),
\eea 
where $(\mathcal{P},\mathcal{F})$ are functions and $(\beta,\omega)$ are one-forms on the four-dimensional flat base space $ds_4^2(\mathcal{B})$. The light-cone coordinates are defined as
\bea 
u=\frac{t-y}{\sqrt{2}},\quad v=\frac{t+y}{\sqrt{2}},
\eea 
with $t$ and $y$ the time and the $S^1$ coordinate. Other field contents of this theory are described in terms of three potential functions $Z_1$, $Z_2$ and $Z_4$ and three magnetic two-form  $\Theta_1$, $\Theta_2$ and $\Theta_4$, in particular $P\equiv Z_1Z_2-Z_4^2$. Their governing equations are organized into a set of linear differential equations with two layers\cite{Bena:2011dd}. The superstrate constructed to date are all superposition of solutions that involve with different single-mode excitation. The single-mode solutions have mode dependence of the form
\bea 
\chi_{k,m,n}\equiv \frac{\sqrt{2}}{R_y}(m+n)v+(k-m)\phi-m\psi,
\eea 
where $\phi$ and $\psi$ are the two angular coordinates of the base space $\mathcal{B}$. The three mode numbers $(k,m,n)$ are constrained by the smoothness condition such that $k$ is a positive integer and $m,n$ are non-negative integers with $m\leq k$. It was proposed in \cite{Bena:2017xbt} that the single-mode solution $(k,m,n)$ is dual to the  CFT state of the form
\bea 
(|++\rangle_1)^{N_1}(\frac{(J_-^+)^m}{m!}\frac{(L_{-1}-J_{-1}^3)^n}{n!}|00\rangle_k)^{N_{k,m,n}},\quad N_1+kN_{k,m,n}=N.
\eea 
Among them the family of solutions with $(k,m,n)=(1,0,n)$ which are dual to \eqref{HeavyOperator} has special properties and hidden symmetries \cite{Bena:2017upb}. The massless scalar wave equation is separable and the six-dimensional metric admits a dimensional reduction to a three-dimensional space time. Below we will concentrate on this integrable family of superstrata.

\subsection{Integrable superstrata}
This family of solution is firstly constructed in \cite{Bena:2017xbt}. The four-dimensional metric $ds_4(\mathcal{B})$ is more conveniently written in the spherical bipolar coordinates
\bea\label{4DMetric}
ds_4^2=\Sigma \left( \frac{dr^2}{r^2+a^2}+d\theta^2\right)+(r^2+a^2)\sin^2\theta d\phi^2+r^2\cos^2\theta d\psi^2,\quad \Sigma\equiv r^2+a^2\cos^2\theta.
\eea 
The remaining parts of the solution are given by \cite{Bena:2017xbt}
\bea
&&Z_1=\frac{Q_1}{\Sigma}+\frac{R_y^2}{2Q_5}b^2\frac{\Delta}{\Sigma}\cos 2\chi,\quad Z_2=\frac{Q_5}{\Sigma},\quad Z_4=bR_y\frac{\Delta}{\Sigma}\cos\chi,\quad  \Delta\equiv\frac{a r^n}{(r^2+a^2)^{\frac{1+n}{2}}}\sin\theta,\nonumber\\
&&\mathcal{F}=-\frac{b^4}{a^2}(1-\frac{r^{2n}}{(r^2+a^2)^n}),\quad \omega=\omega_0+\frac{R_yb^2}{\sqrt{2}\Sigma}(1-\frac{r^{2n}}{(r^2+a^2)^n})\sin^2\theta d\phi,\\
&&\beta=\frac{R_y a^2}{\sqrt{2}\Sigma}(\sin^2\theta d\phi-\cos^2\theta d\psi),\quad \omega_0=\frac{R_y a^2}{\sqrt{2}\Sigma}(\sin^2\theta d\phi+\cos^2\theta d\psi),\quad \frac{Q_1Q_5}{R_y^2}=a^2+\frac{b^2}{2}.\nonumber
\eea 
This solution describes the coherent superposition of state \eqref{HeavyOperator} with average numbers of $|++\rangle_1$ and $|00\rangle_1$ strands $N R_y^2a^2/(Q_1Q_5)$ and  $N R_y^2b^2/(2Q_1Q_5)$ respectively. Introducing the $AdS_3$ coordinates $(x^1,x^2,x^3)\equiv (r,t,y)$ and the $S^3$ coordinates $(y^1,y^2,y^3)=(\theta,\phi,\psi)$ one can recast the six-dimensional metric in the form
\bea\label{Fibration}
ds_6^2=V^{-2}g_{\mu\nu}dx^\mu dx^\nu+G_{\alpha\beta}(dy^\alpha+A^\alpha_\mu dx^\mu)(dy^\beta+A_\nu^\beta dx^\nu),\quad V^2\equiv\frac{\text{det}G}{\text{det}G|_{r\rightarrow \infty}}.
\eea 
One finds that the three-dimensional metric $g_{\mu\nu}$ does not depend on any $S^3$ coordinates so that we can perform a dimensional reduction on $S^3$. After the dimensional reduction the resulting three-dimensional metric is \cite{Bena:2017upb}
\bea \label{ReducedMetric}
\frac{g_{\mu\nu}}{\sqrt{Q_1Q_5}}=&& \frac{F_2(r)dr^2}{r^2+a^2}+\frac{2a^2r^2(r^2+a^2)F_2(r)}{F_1(r)R_y^2}du^2 \nonumber \\
&&-\frac{2F_1(r)}{a^2(2a^2+b^2)^2 R_y^2}(dv+\frac{a^2(a^4+(2a^2+b^2)r^2)}{F_1(r)}du)^2,
\eea 
where
\bea 
&& F_0(r)=1-\frac{r^{2n}}{(r^2+a^2)^n},\quad F_1(r)=a^6-b^2(2a^2+b^2)r^2 F_0(r),\nonumber \\
&&F_2(r)=1-\frac{a^2 b^2}{(2a^2+b^2)}\frac{r^{2n}}{(r^2+a^2)^{n+1}}.
\eea 
As a consequence, the six-dimensional scalar Laplacian operator with respect to $ds_6^2$ gets simplified into a three-dimensional scalar Laplacian operator with respect to $g_{\mu\nu}$. In the following we will use this simplification to compute the four-point correlators. 

\subsection{Holographic bosonic four-point correlators}

In the CFT result \eqref{Correlator1}, there is a term of order $N$ which corresponds to the disconnected part of the holographic correlator. Therefore this term is given by the modulus square of the three-point function which describes the expectation values of the light operator in the non-trivial background. Since the bosonic light operator is dual to a minimally coupled scalar, its expectation value vanishes \footnote{For the fermionic correlator, the holographic calculation of this disconnected part can be found in \cite{Moscato:2017usq}.}. This is also reflected in the Ward identity \eqref{WardIdentity}.  Below we will focus on the connected part.

The holographic computation of the correlator is involved in solving the six-dimensional Laplacian equation
\bea\label{6DLaplace}
\square_6 B=0,
\eea 
with the boundary condition
\bea\label{Boundary}
B\sim \delta(t,y)+\frac{ \mathbf{b}(t,y)}{r^2},\quad r\rightarrow \infty.
\eea 
As mentioned above, the six-dimensional Laplacian operator can be simplified to the three-dimensional one defined with the reduced metric \eqref{ReducedMetric}. The four-point  function can be exacted from the function $\mathbf{b}(t,y)$ through \cite{Galliani:2017jlg}
\bea
\langle O_H(0)\bar{O}(\infty)O_L(1,1)\bar{O}_L(z,\bar{z})\rangle=\frac{1}{|1-z|^4}\mathcal{G}^{bos}(z,\bar{z})=(z\bar{z})^{-1}\mathbf{b}(z,\bar{z}),
\eea 
where 
\bea 
z=e^{i\frac{t+y}{R_y}}=e^{\frac{t_e+iy}{R_y}},\quad \bar{z}=e^{i\frac{t-y}{R_y}}=e^{\frac{t_e-iy}{R_y}},\quad \text{with }t_e\equiv it\text{ the Eucliean time}. 
\eea 
Even though the Laplacian equation \eqref{6DLaplace} is separable, the  radial  differential equation is still very hard to solve. Therefore we will adopt an approximation scheme that was used in \cite{Galliani:2017jlg}. Taking $b$ as a small quantity we perform the $b$ expansion keeping $Q_1$, $Q_5$ and $R$ and hence $a_0\equiv a^2+b^2/2$ fixed:
\bea\label{Expansion}
&& B=B_0+b^2 B_1,\quad \square_3=\square_0+b^2 \square_1,\nonumber \\
&& \square_0 B_0=0,\quad \square_0 B_1=-\square_1 B_0\equiv J.
\eea
The zero order of the background is global $AdS_3$ and the first order metric is shown in Appendix \ref{FirstOrderMetric}. The boundary condition \eqref{Boundary} implies that $B_0$ is the bulk-to-boundary propagator of dimension $\Delta=2$ in global  $AdS_3$:
\bea\label{BulkToB}
B_0(r,t,y)=K_2(r,t,y|t'=0,y'=0)=\left[   \frac{1}{2}\frac{a_0}{\sqrt{r^2+a_0^2}\cos(t/R)-r \cos(y/R)}   \right]^2.
\eea 
The information of the first order correction of the correlator is encoded in this integral 
\bea
\mathbf{b}_1(t,y)=-\frac{b^2}{2\pi} \int d^3 r'\sqrt{-g}K_2(t',y',r'|t,y)J(t',y',r').
\eea
Substituting \eqref{BulkToB} into \eqref{Expansion} gives the expression of the source current
\bea
&&J=-\frac{r}{a_0^2+r^2}\p_r B_0 +\frac{a_0^2 R^2}{(a_0^2+r^2)^2}\p^2_tB_0-\frac{1}{2}D_n (\p_{t}-\p_y)^2B_0,\\ 
&&D_n\equiv\frac{R^2}{a_0^2+r^2}[(\frac{r^2}{(a_0^2+r^2)})^n-1]=\frac{R^2}{a_0^2}\sum_{k\geq 1}^{n} \begin{pmatrix}
	n  \\
	k \\
\end{pmatrix} (-1)^k (\frac{a_0^2}{a_0^2+r^2})^{k+1}. \nonumber
\eea
When $n=0$, one can find that the first two terms are exactly the source current obtained in \cite{Galliani:2017jlg} on a $1/4$ BPS geometry. The new contribution is
\bea
\frac{b^2}{2\pi} \frac{R^2}{2a_0^2}\sum_{k\geq 2}^{n+1} \begin{pmatrix}
	n  \\
	k-1 \\
\end{pmatrix} (-1)^k\int d^3 r' \sqrt{-g}K_2(\mathbf{r_e}'|t,y)(\p_{t'_e}+i\p_{y'})^2K_2(\mathbf{r_e}'|0) (\frac{a_0^2}{a_0^2+r^2})^{k},
\eea
where we have replaced $t$ with $-it_e$ and $k$ with $k-1$. Omitting the factors the integral can be computed by integration by parts twice
\bea
I_k&=&\int d^3 r' \sqrt{-g}K_2(\mathbf{r_e}'|t,y)(\p_{t'_e}+i\p_{y'})^2K_2(\mathbf{r_e}'|0) (\frac{a_0^2}{a_0^2+r^2})^{k}\nonumber \\
&=&(\p_{t_e}+i\p_y)^2\int d^3 r' \sqrt{-g}K_2(\mathbf{r_e}'|t,y)K_2(\mathbf{r_e}'|0) (\frac{a_0^2}{a_0^2+r^2})^{k}\\
&=&(\p_{t_e}+i\p_y)^2\int d^3 r' \sqrt{-g}K_2(\mathbf{r_e}'|t,y)K_2(\mathbf{r_e}'|0)K_k(\mathbf{r_e}'|\infty)K_k(\mathbf{r_e}'|-\infty)\nonumber\\
&=&(\p_{t_e}+i\p_y)^2 (|z|^2\hat{D}_{kk22})=\left(\frac{2}{R}\bar{z}\p_{\bar{z}}\right)^2 (|z|^2\hat{D}_{kk22}),\nonumber
\eea
where $\hat{D}$ is the regularized D-function.
Adding the trivial contribution $1/|1-z|^4$ and the $1/4$ BPS contribution \cite{Galliani:2017jlg}, one finds the holographic bosonic correlator
\bea\label{HolographicCorrelator}
\frac{1}{|1-z|^4}\mathcal{G}^{bos}(z,\bar{z})&=&\frac{1}{|1-z|^4}+\frac{b^2}{\pi a_0^2}\p\bar{\p}\left[-\frac{\pi}{2}\frac{1}{|1-z|^2}+|z|^2\hat{D}_{1122}\right]\nonumber \\
&&\quad +\frac{b^2}{\pi a_0^2|z|^2}\sum_{k\geq 2}^{n+1} \begin{pmatrix}
	n  \\
	k-1 \\
\end{pmatrix} (-1)^k (\bar{z}\bar{\p})^2(|z|^2\hat{D}_{kk22}).
\eea 
Because of the last term with $(\bar{z}\bar{\p})^2$, this bosonic correlator can not be simply converted to fermionic one via the Ward identity \eqref{WardIdentity}. However it is straightforward to obtain the bosonic correlators from the fermionic ones. For example, the $n=1$ fermionic correlator \eqref{Correlator1} gives this simple result
\bea 
\mathcal{G}^{bos}_{n=1,CFT}=1-\frac{B^2}{2N}+\frac{B^2}{2N}z(2-z).
\eea 
As a comparison, from holographic calculation \eqref{HolographicCorrelator} we can derive this correlator in the strong coupling regime
\bea 
\mathcal{G}^{bos}_{n=1,gravity}=1-\frac{b^2}{2a_0^2}+|1-z|^4 \frac{b^2}{\pi a_0^2}\left[\p\bar{\p}(|z|^2\hat{D}_{1122})+\frac{1}{|z|^2}(\bar{z}\bar{\p})^2(|z|^2\hat{D}_{2222})\right].
\eea 
With the parameter identification $b=B$ and $a_0^2=N$, we find that only the $z$--independent term matches. So the correlator is not protected as expected.

\section{Conclusion}

In this paper we extended the calculations of HHLL correlators \eqref{Correlator} in the context of the $AdS_3/CFT_2$ correspondence from the $1/4$-BPS sector to the $1/8$-BPS sector . The field theory results \eqref{Correlator1}, \eqref{CorrelatorN} and \eqref{DiagonalN} are computed at the free orbifold point of the D1-D5 CFT. The gravity results \eqref{HolographicCorrelator} are computed perturbatively on an integrable family of three-charge microstate geometries. 

Our calculation does not rely on the standard Witten diagram method. By comparing the two results we find that this correlators are not protected as their $1/4$ BPS cousins. The mismatch is due to the exchange of non-chiral operators. These results 
describe correlators in the weak and strong coupling limits respectively. However it is possible to construct full $1/8$ BPS four point functions following a strategy which is suggested in \cite{Giusto:2018ovt}. Several explicit examples have been remarkably obtained in \cite{Bombini:2019vnc}. One of the most important information holographic correlators is the spectrum of the double-trace operators at strong coupling. Our results can shed light on the calculation of anomalous dimensions of double-trace operators as in the case of $AdS_5/CFT_4$ duality.

 In contrast to the two-charge microstate geometries, three-charge ones describe non-degenerate black holes. Therefore our results would be used to study general mechanism for information conservation of a realistic black hole.


\section*{Acknowledgments}
We thank Alessandro Bombini and Andrea Galliani for comments. The work was in part supported by NSFC Grant No.~11275010, No.~11335012, No.~11325522 and No. 11735001.\appendix

\section{Some details on correlators with generic $n$}
\label{GeneralN}
\renewcommand{\theequation}{A.\arabic{equation}}
\setcounter{equation}{0}
In this appendix we provide the details of the deviation of \eqref{GeneralOff} and \eqref{GenralDiagonalA}.

First let's compute the off-diagonal contribution:
\bea 
&&\langle ++|O_3 (L_{-1}-J_{-1}^3)^n |00\rangle =\langle ++|[O_3,M_-],M_-],\dots,M_-]|00\rangle.
\eea 
Using the commutation relations \eqref{Commutation2}
\bea
&[O_3,M_-]&=(-\p+\frac{1}{2}z^{-1})O_3, \nonumber \\
&[[O_3,M_-],M_-]&=-\p (-\p) O_3-\frac{1}{2}z^{-1}\p O_3-[J_{-1}^3,\p O_3]+\frac{1}{2}z^{-1}[J_{-1}^3,O_3] \nonumber \\
&& = \p^2 O_3 -\frac{1}{2}z^{-1}\p O_3-\p (\frac{1}{2}z^{-1}O_3)+\frac{1}{4}z^{-2}O_3=(-\p+\frac{1}{2}z^{-1})^2 O_3,\nonumber
\eea  
one can obtain
\bea 
\langle ++|O_3 (L_{-1}-J_{-1}^3)^n |00\rangle=(-\p+\frac{1}{2}z_3^{-1})^n z_3^{-1/2}\bar{z}_3^{-1/2},
\eea 
\bea 
\langle 00|(L_{+1}-J_{+1}^3)^n O_4|++\rangle =  (z_4^2 \p+\frac{3}{2}z_4)^nz_4^{-1/2}\bar{z}_4^{-1/2}.
\eea 
Comparing with the previous results, we have
\bea 
\mathcal{G}_{off}(z,\bar{z})=z^n\frac{|1-z|^2}{|z|}\frac{|A|^2|B|^2(N-1)}{N^2}.
\eea 
The most cumbersome term is 
\bea 
\langle 00|(L_{+1}-J_{+1}^3)^nO_3  O_4 (L_{-1}-J_{-1}^3)^n|00\rangle \equiv \langle M_+^n O_3 O_4 M_-^n\rangle \equiv G_n.
\eea 
Let us derive a recursion relationship. Introduce the convenient notations as below
\bea
&&G_{n+1}= \langle M_+^n M_+O_3 O_4 M_- M_-^n\rangle= \langle M_+^n M_+ M_-O_3 O_4M_-^n\rangle-(D_-^3+D_-^4) B_{n}, \nonumber \\
&&B_{n}= \langle M_+^n M_+ O_3 O_4M_-^n\rangle.
\eea 
The  first term is given by
\bea 
&& \langle M_+^n M_+ M_-O_3 O_4M_-^n\rangle=\langle M_-M_+^n M_+O_3O_4M_-^n\rangle-\sum\limits_{p=0}^n \langle M_+^p[M_-,M_+]M_+^{n-p} O_3 O_4M_-^n\rangle \nonumber \\
&&\quad\qquad= 0+\sum\limits_{p=0}^n \langle M_+^p(1+2L_0)M_+^{n-p} O_3 O_4M_-^n\rangle =\sum\limits_{p=0}^n(1+2p)G_n=(n+1)^2G_n. \nonumber
\eea 
To compute $B_n$ we move one $M_+$ to the left
\bea 
B_{n}&=& (D_+^3+D_+^4)G_n+\langle M_+^n O_3 O_4M_+M_-^n\rangle=(D_+^3+D_+^4)G_n+n^2 B_{n-1}. \nonumber
\eea 
In summary, we have derived a coupled recurrence equation for the correlator
\bea 
&&G_{n+1}=(n+1)^2G_n-(D_-^3+D_-^4) B_{n}, \\
&&B_{n+1}=(n+1)^2B_n+(D_+^3+D_+^4) G_{n+1},\\
&&G_{0}=\langle O_3 O_4\rangle,\quad B_{0}=(D_+^3+D_+^4) \langle O_3 O_4\rangle.
\eea 
Canceling $B_n$ we can get recurrence equation solely with $G_n$
\bea 
&&G_{n+2}=[(n+1)^2+(n+2)^2-D]G_{n+1}-(n+1)^4 G_{n},\\
&&G_{1}=(1-D)G_{0},\quad D\equiv D_-D_+.
\eea 
After computing the first few terms explicitly, we find the expression gets more and more involved. Below we derive an general expression of $G_n$ in another way.
Define
\begin{align}
F_k^n \equiv M _ { + } ^ { k } O _ { 3 } O _ { 4 } M _ { - } ^ { n }
\end{align}
then
\bea
G_0=\langle F^0_0 \rangle=\langle O_3 O_4\rangle,\hspace{3ex}
G_n=\langle F^n_n \rangle.
\eea
It is easy to show that
\bea
F^n_0= O _ { 3 } O _ { 4 } M _ { - } ^ { n }=(M_- - D_-)^n F^0_0. \label{eq:291}
\eea
Now, we will evaluate $F^n_k$. Firstly, we evaluate it when $n=1,2,3$.
\be\begin{split}
	F_1^n
	&= M _ { + }  O _ { 3 } O _ { 4 } M _ { - } ^ { n }\\
	&= D_+ O _ { 3 } O _ { 4 } M _ { - } ^ { n }+O _ { 3 } O _ { 4 } \left[M^n_- M_+ -\sum_{p=0}^{n-1} M^{n-1-p}_- [M_-,M_+]M^p_- \right]\\
	&=D_+ F^n_0+O _ { 3 } O _ { 4 } \left[M^n_- M_+ + \sum_{p=0}^{n-1} M^{n-1-p}_- (1+2 L_0 -2J_0)M^p_- \right]\\
	&=D_+ F^n_0+O _ { 3 } O _ { 4 } M^{n-1}_- \left[M_-M_+ + n(n+2 L_0 -2J_0)\right]\\
	&=D_+ F^n_0+n^2 F^{n-1}_0 +...\\
\end{split}\ee
where ... represents the terms which annihilate the vacuum, that is to say, $...|00\rangle=0$ . What we want is the vacuum expectation value of $F^n_k=M_+^k F^n_0$, so we will neglect ... and write $F^n_1$ as
\be
F_1^n=D_+ F^n_0+n^2 F^{n-1}_0.
\ee
Plugging \eqref{eq:291} into the last equation we get
\be\begin{split}
	F_1^n&=D_+ F^n_0+n^2 F^{n-1}_0\\
	&=\left(D_+(M_- - D_-)^n+n^2(M_- - D_-)^{n-1}\right)F^{0}_0\\
	&\equiv \bar{M}_n F^{0}_0
\end{split}\ee
Notice that $\langle00| \bar{M}_n=\langle00|\left(D_+(-D_-)^n+n^2(-D_-)^{n-1}\right)$.
\be\begin{split}
	F_2^n&=M_+F^n_1=M_+\left(D_+ F^n_0+n^2 F^{n-1}_0\right)\\
	&=D_+ F^n_1+n^2 F^{n-1}_1\\
	&=\left(D_+ \bar{M}_n+n^2\bar{M}_{n-1}\right)F^0_0
\end{split}\ee
\be\begin{split}
	F_3^n=\left(D_+^2 \bar{M}_n+2D_+\bar{M}_{n-1}n^2+n^2(n-1)^2\bar{M}_{n-2}\right)F^0_0
\end{split}\ee
Hence, we find that
\be\label{Fnk}\begin{split}
	F_{k+1}^n=\sum_{i=0}^{k}C_k^i \frac{n!^2}{(n-i)!^2}D_+^{k-i}\bar{M}_{n-i}F^0_0, \quad C_k^i\equiv \begin{pmatrix}
		k\\
		i \\
	\end{pmatrix},\quad \forall k\geq 0.
\end{split}\ee
This result \eqref{Fnk} can be shown by the mathematical induction with the help of the identity $C_{k+1}^i=C_k^i+C_{k}^{i-1}$. Therefore, the general expression of $G_n$ can be written as
\be\begin{split}\label{DiagonalN}
	G_n&=\langle F_{n}^n\rangle=\sum_{i=0}^{n-1}C_{n-1}^i \frac{n!^2}{(n-i)!^2}D_+^{n-1-i}\langle \bar{M}_{n-i}F^0_0\rangle\\
	&=\sum_{i=0}^{n-1}C_{n-1}^i \frac{n!^2}{(n-i)!^2}D_+^{n-1-i}(-1)^{n-i}\left(D_+D_-^{n-i}-(n-i)^2D_-^{n-1-i}\right)G_0.
\end{split}\ee

\section{Reduced metric to the order $b^2$}
\label{FirstOrderMetric}
In this appendix, we present the reduced metric to the leading order of $b^2$:
\bea 
&&g_{tt}=-\frac{a_0^2+r^2}{a_0^2 R_y^2}+\frac{b^2}{2a_0^4 R_y^2}(2a_0^2+r^2-\frac{r^{2n+2}}{(a_0^2+r^2)^n}),\nonumber \\
&&g_{ty}=\frac{b^2}{2a_0^4R_y^2}(r^2-\frac{r^{2n+2}}{(a_0^2+r^2)^n}),\\
&&g_{yy}=\frac{r^2}{a_0^2 R_y^2}+\frac{b^2}{2a_0^4R_y^2}(r^2-\frac{r^{2n+2}}{(a_0^2+r^2)^n}),\nonumber\\
&&g_{rr}=\frac{1}{a_0^2+r^2}+\frac{b^2}{2}(\frac{1}{(a_0^2+r^2)^2}-\frac{r^{2n}}{(a_0^2+r^2)^{n+2}}).\nonumber
\eea


\newpage
\begin{thebibliography}{30}

\bibitem{Maldacena:1997re}
  J.~M.~Maldacena,
  ``The Large $N$ Limit of Superconformal Field Theories and Supergravity,''
  Int.\ J.\ Theor.\ Phys.\  {\bf 38} (1999) 1113
   [Adv.\ Theor.\ Math.\ Phys.\  {\bf 2} (1998) 231]
  doi:10.1023/A:1026654312961, 10.4310/ATMP.1998.v2.n2.a1
  [hep-th/9711200].
 
 
\bibitem{Gubser:1998bc}
  S.~S.~Gubser, I.~R.~Klebanov and A.~M.~Polyakov,
  ``Gauge Theory Correlators from Noncritical String Theory,''
  Phys.\ Lett.\ B {\bf 428} (1998) 105
  doi:10.1016/S0370-2693(98)00377-3
  [hep-th/9802109].
  

\bibitem{Witten:1998qj}
  E.~Witten,
  ``Anti-de~Sitter Space and Holography,''
  Adv.\ Theor.\ Math.\ Phys.\  {\bf 2} (1998) 253
  doi:10.4310/ATMP.1998.v2.n2.a2
  [hep-th/9802150].

  
\bibitem{DHoker:1999mqo}
E.~D'Hoker, D.~Z.~Freedman and L.~Rastelli,
``AdS / CFT Four Point Functions: How to Succeed at Z Integrals without Really Trying,''
Nucl.\ Phys.\ B {\bf 562} (1999) 395
doi:10.1016/S0550-3213(99)00526-X
[hep-th/9905049];\\

G.~Arutyunov, A.~Pankiewicz and S.~Theisen,
``Cubic Couplings in D = 6 ${\mathcal{N}}\!=4$B Supergravity on Ad$S^3$  $\times\ S^3$ ,''
Phys.\ Rev.\ D {\bf 63} (2001) 044024
doi:10.1103/PhysRevD.63.044024
[hep-th/0007061].


\bibitem{Galliani:2017jlg} 
A.~Galliani, S.~Giusto and R.~Russo,
``Holographic 4-point correlators with heavy states,''
JHEP {\bf 1710}, 040 (2017)
doi:10.1007/JHEP10(2017)040
[arXiv:1705.09250 [hep-th]].\\
 A.~Bombini, A.~Galliani, S.~Giusto, E.~Moscato and R.~Russo,
  ``Unitary 4-Point Correlators from Classical Geometries,''
  Eur.\ Phys.\ J.\ C {\bf 78} (2018) no.1,  8
  doi:10.1140/epjc/s10052-017-5492-3
  [arXiv:1710.06820 [hep-th]].
  
\bibitem{Galliani:2016cai}
  A.~Galliani, S.~Giusto, E.~Moscato and R.~Russo,
  ``Correlators at Large C without Information Loss,''
  JHEP {\bf 1609} (2016) 065
  doi:10.1007/JHEP09(2016)065
  [arXiv:1606.01119 [hep-th]].
\bibitem{Giusto:2015dfa}
  S.~Giusto, E.~Moscato and R.~Russo,
  ``AdS$_{3}$ Holography for 1/4 and 1/8 BPS Geometries,''
  JHEP {\bf 1511} (2015) 004
  doi:10.1007/JHEP11(2015)004
  [arXiv:1507.00945 [hep-th]].

\bibitem{Giusto:2018ovt}
  S.~Giusto, R.~Russo and C.~Wen,
  ``Holographic Correlators in AdS$_3$,''
  arXiv:1812.06479 [hep-th].






\bibitem{Balasubramanian:2005qu}
  V.~Balasubramanian, P.~Kraus and M.~Shigemori,
  ``Massless Black Holes and Black Rings as Effective Geometries of the D1-D5 System,''
  Class.\ Quant.\ Grav.\  {\bf 22} (2005) 4803
  doi:10.1088/0264-9381/22/22/010
  [hep-th/0508110].


\bibitem{Lunin:2012gz}
  O.~Lunin and S.~D.~Mathur,
  ``A Toy Black Hole S-Matrix in the D1-D5 CFT,''
  JHEP {\bf 1302} (2013) 083
  doi:10.1007/JHEP02(2013)083
  [arXiv:1211.5830 [hep-th]].


\bibitem{Balasubramanian:2016ids}
  V.~Balasubramanian, B.~Craps, B.~Czech and G.~Ssrosi,
  ``Echoes of Chaos from String Theory Black Holes,''
  JHEP {\bf 1703} (2017) 154
  doi:10.1007/JHEP03(2017)154
  [arXiv:1612.04334 [hep-th]].

\bibitem{Fitzpatrick:2014vua}
  A.~L.~Fitzpatrick, J.~Kaplan and M.~T.~Walters,
  ``Universality of Long-Distance AdS Physics from the CFT Bootstrap,''
  JHEP {\bf 1408} (2014) 145
  doi:10.1007/JHEP08(2014)145
  [arXiv:1403.6829 [hep-th]].\\
  A.~L.~Fitzpatrick, J.~Kaplan and M.~T.~Walters,
  ``Virasoro Conformal Blocks and Thermality from Classical Background Fields,''
  JHEP {\bf 1511} (2015) 200
  doi:10.1007/JHEP11(2015)200
  [arXiv:1501.05315 [hep-th]].\\
  A.~L.~Fitzpatrick, J.~Kaplan and M.~T.~Walters,
  ``Virasoro Conformal Blocks and Thermality from Classical Background Fields,''
  JHEP {\bf 1511} (2015) 200
  doi:10.1007/JHEP11(2015)200
  [arXiv:1501.05315 [hep-th]].


\bibitem{Hijano:2015zsa}
  E.~Hijano, P.~Kraus, E.~Perlmutter and R.~Snively,
  ``Witten Diagrams Revisited: the AdS Geometry of Conformal Blocks,''
  JHEP {\bf 1601} (2016) 146
  doi:10.1007/JHEP01(2016)146
  [arXiv:1508.00501 [hep-th]].


\bibitem{Hijano:2015qja}
  E.~Hijano, P.~Kraus, E.~Perlmutter and R.~Snively,
  ``Semiclassical Virasoro Blocks from AdS$_{3}$ Gravity,''
  JHEP {\bf 1512} (2015) 077
  doi:10.1007/JHEP12(2015)077
  [arXiv:1508.04987 [hep-th]].

\bibitem{daCunha:2016crm} 
B.~Carneiro da Cunha and M.~Guica,
``Exploring the BTZ bulk with boundary conformal blocks,''
arXiv:1604.07383 [hep-th].
\bibitem{Lunin:2001jy}
  O.~Lunin and S.~D.~Mathur,
  ``AdS / CFT Duality and the Black Hole Information Paradox,''
  Nucl.\ Phys.\ B {\bf 623} (2002) 342
  doi:10.1016/S0550-3213(01)00620-4
  [hep-th/0109154].
\bibitem{Lunin:2002iz}
  O.~Lunin, J.~M.~Maldacena and L.~Maoz,
  ``Gravity Solutions for the D1-D5 System with Angular Momentum,''
  hep-th/0212210.

\bibitem{Kanitscheider:2007wq}
  I.~Kanitscheider, K.~Skenderis and M.~Taylor,
  ``Fuzzballs with Internal Excitations,''
  JHEP {\bf 0706} (2007) 056
  doi:10.1088/1126-6708/2007/06/056
  [arXiv:0704.0690 [hep-th]].


\bibitem{Bena:2015bea}
  I.~Bena, S.~Giusto, R.~Russo, M.~Shigemori and N.~P.~Warner,
  ``Habemus Superstratum! a Constructive Proof of the Existence of Superstrata,''
  JHEP {\bf 1505} (2015) 110
  doi:10.1007/JHEP05(2015)110
  [arXiv:1503.01463 [hep-th]].
\bibitem{Bena:2017xbt}
  I.~Bena, S.~Giusto, E.~J.~Martinec, R.~Russo, M.~Shigemori, D.~Turton and N.~P.~Warner,
  ``Asymptotically-Flat Supergravity Solutions Deep Inside the Black-Hole Regime,''
  JHEP {\bf 1802} (2018) 014
  doi:10.1007/JHEP02(2018)014
  [arXiv:1711.10474 [hep-th]].\\
   I.~Bena, S.~Giusto, E.~J.~Martinec, R.~Russo, M.~Shigemori, D.~Turton and N.~P.~Warner,
  Phys.\ Rev.\ Lett.\  {\bf 117}, no. 20, 201601 (2016)
  doi:10.1103/PhysRevLett.117.201601
  [arXiv:1607.03908 [hep-th]].
\bibitem{Tian:2016ucg}
  W.~Tian,
  ``Multicenter Superstrata,''
  Phys.\ Rev.\ D {\bf 94} (2016) no.6,  066011
  doi:10.1103/PhysRevD.94.066011
  [arXiv:1607.08884 [hep-th]].
\bibitem{Ceplak:2018pws}
  N.~Ceplak, R.~Russo and M.~Shigemori,
  ``Supercharging Superstrata,''
  arXiv:1812.08761 [hep-th].
  
  \bibitem{Bakhshaei:2018vux} 
  E.~Bakhshaei and A.~Bombini,
  ``Three-charge superstrata with internal excitations,''
  Class.\ Quant.\ Grav.\  {\bf 36}, no. 5, 055001 (2019)
  doi:10.1088/1361-6382/ab01bc
  [arXiv:1811.00067 [hep-th]].
 
  \bibitem{Heidmann:2019zws} 
  P.~Heidmann and N.~P.~Warner,
  ``Superstratum Symbiosis,''
  arXiv:1903.07631 [hep-th].
  \bibitem{Bombini:2017got} 
  A.~Bombini and S.~Giusto,
  ``Non-extremal superdescendants of the D1D5 CFT,''
  JHEP {\bf 1710}, 023 (2017)
  doi:10.1007/JHEP10(2017)023
  [arXiv:1706.09761 [hep-th]].
\bibitem{Bena:2017upb}
  I.~Bena, D.~Turton, R.~Walker and N.~P.~Warner,
  ``Integrability and Black-Hole Microstate Geometries,''
  JHEP {\bf 1711} (2017) 021
  doi:10.1007/JHEP11(2017)021
  [arXiv:1709.01107 [hep-th]].
 
 \bibitem{Bombini:2019vnc} 
  A.~Bombini and A.~Galliani,
  ``AdS$_3$ four-point functions from $\frac{1}{8}$-BPS states,''
  arXiv:1904.02656 [hep-th].
   
 \bibitem{Lunin:2012gp} 
 O.~Lunin, S.~D.~Mathur and D.~Turton,
 ``Adding momentum to supersymmetric geometries,''
 Nucl.\ Phys.\ B {\bf 868}, 383 (2013)
 doi:10.1016/j.nuclphysb.2012.11.017
 [arXiv:1208.1770 [hep-th]].
 
 \bibitem{Bena:2008wt}
 I.~Bena, N.~Bobev and N.~P.~Warner,
 ``Spectral Flow, and the Spectrum of Multi-Center Solutions,''
 Phys.\ Rev.\ D {\bf 77} (2008) 125025
 doi:10.1103/PhysRevD.77.125025
 [arXiv:0803.1203 [hep-th]].
 
 \bibitem{Giusto:2013rxa}
 S.~Giusto, L.~Martucci, M.~Petrini and R.~Russo,
 ``6D microstate geometries from 10D structures,''
 Nucl.\ Phys.\ B {\bf 876} (2013) 509
 doi:10.1016/j.nuclphysb.2013.08.018
 [arXiv:1306.1745 [hep-th]].\\
 S.~Giusto and R.~Russo,
 ``Superdescendants of the D1D5 CFT and their dual 3-charge geometries,''
 JHEP {\bf 1403}, 007 (2014)
 doi:10.1007/JHEP03(2014)007
 [arXiv:1311.5536 [hep-th]].
 
  \bibitem{Gutowski:2003rg} 
  J.~B.~Gutowski, D.~Martelli and H.~S.~Reall,
  ``All Supersymmetric solutions of minimal supergravity in six- dimensions,''
  Class.\ Quant.\ Grav.\  {\bf 20}, 5049 (2003)
  doi:10.1088/0264-9381/20/23/008
  [hep-th/0306235].
  \bibitem{Bena:2011dd} 
  I.~Bena, S.~Giusto, M.~Shigemori and N.~P.~Warner,
  ``Supersymmetric Solutions in Six Dimensions: A Linear Structure,''
  JHEP {\bf 1203}, 084 (2012)
  doi:10.1007/JHEP03(2012)084
  [arXiv:1110.2781 [hep-th]].
\bibitem{Moscato:2017usq} 
  E.~Moscato,
  ``Black Hole Microstates And Holography in the D1D5 CFT''. 
  












\end{thebibliography}
\end{document}